\documentclass[11pt,a4paper]{article}
\usepackage[utf8]{inputenc}
\usepackage[T1]{fontenc}
\usepackage[english]{babel}
\usepackage{amsmath,amsthm}
\usepackage{amsfonts}
\usepackage{amssymb}
\usepackage{graphicx}
\usepackage{subfig}
\usepackage[left=2cm,right=2cm,top=2cm,bottom=2cm]{geometry}
\usepackage{authblk}
\usepackage{xcolor}

\title{Long term surface-air temperature variations over the Indian region during 1970-2009}

\author[$\dagger1$]{Kamsali Nagaraja}
\author[]{S. C. Chakravarty}

\affil[]{Department of Physics, Bangalore University, Bengaluru 560056 India}
\affil[$\dagger$]{Corresponding author: kamsalinagaraj@bub.ernet.in}
\date{}

\begin{document}

\maketitle

\begin{abstract}
Global and regional annual mean temperature data have been analysed by many groups to determine the linear trends of temperature over climatological time scales. The near consistent results generally show an increase of about 0.07$^\circ$C per decade during the 20th Century. But many basic questions including spatial and temporal data gaps, non-uniform distribution of observing sites, superposition of internal/natural variations at different scales with parametric feedbacks etc., still remain unresolved. This paper mainly deals with a detailed study of the climatological variations of surface-air temperatures over the Indian region using a well-tested, verified, researched and gridded (1$^\circ$x1$^\circ$) daily mean temperature data set for the period 1970-2009. The annual mean temperatures estimated with different spatial integration show linear trends with an increase of about 0.4 $^\circ$C during this period. A detailed error analysis of the voluminous data shows that the statistical errors at 95\% confidence interval are lower than that of the increase in temperatures determined from its linear trend. The annual mean temperature time series also shows consistent and nearly phase coherent periodic variations of 3-5 years with an average magnitude of $\pm$0.4 $^\circ$C.  Possible causes of these temperature structures due to changes in solar activity, Galactic Cosmic Ray (GCR) fluxes and El Nino and Southern (ENSO) are examined.
\end{abstract}


\section{Introduction}
Global warming and climatological studies have been carried out by subjecting various global and regional data sets to statistical analysis invoking data quality filters, homogenisation, and weighting algorithms to generate annual mean temperature time series in latitude-longitude grids of 5$^\circ$x5$^\circ$ or even of smaller sizes (Jones, 2016, Osborn et al, 2017). The global temperature data sets so treated in all such analyses generally lack in ensuring unbroken spatial-temporal coverage and uniform quality control due to the absence of following standard procedures in conducting observations, recording and archival. At regional scales, the datasets are better controlled in terms of accuracy, resolution and reliability (Feng et al, 2006, Nathaniel et al, 2014). Robust statistical analysis tools have been employed to treat the original regional data for incorporating various methods such as Shepard’s angular distance weighting algorithm (Srivastava et al, 2009), filtering of outliners and error estimation. The multiplicity of approach and multiple pathways to attain the same objective of finding a warming signal from many long term data sets have provided a fairly consistent and acceptable result of a global temperature rise of about 0.07 $^\circ$C per decade during the Twentieth Century (Jones and Moberg, 2003). 

While the scenario of the relative magnitude of global warming is internationally acceptable (IPCC, 2013), there are still concerns that the warming is not continuous and that the recent hiatus (e.g., 1998-2012) of global warming is attributable to long term inherent natural oscillations in mean temperatures obscuring the warming due to anthropogenic forcing (Lin and Franzke, 2015). The cause of scepticism also stems from the fact that the warming is not found to be globally consistent and uniform in spite of the fact that long term data from a small number of evenly-spaced observing stations is all that is required to arrive at a good estimate of global-mean temperature change (Jones and Hulme, 1997), with uncontested confidence levels.  Efforts have been directed to generate equally-spaced globally gridded monthly and daily temperature data sets (New et al, 2000; Caesar et al, 2006) for further advancement in climate research. The coverage of data shown in world maps by Caesar et al (2006) brings out glaring data gaps over the Indian, African, Mexican and South American regions. However, this has not been a major impediment in getting consistent results on global temperature rise and its latitudinal, seasonal and extreme weather anomalies (Parker and Alexander, 2002; Brohan et al, 2006; Safari, 2012). 

The hallmark of global warming/climate change studies has been to select gridded mean temperature data usually at annual or at best seasonal (both temporal) and ~5$^\circ$x5$^\circ$ spatial scales and use trend analysis or linear regression method to estimate the change over a period of a few decades to a century. As mentioned, the results have shown a good degree of consistency in spite of major regional data gaps or inhomogeneity. According to Mudelesee (2010), a possible view of climate change is a time dependent random variable that is composed of trend, outliners/extremes and variability/noise. Along with the trend analysis, there is a need for a more detailed study of the inherent/internal variations which are normally filtered out as much as possible to get a clearer picture on the linear trend of global warming. While this in itself continues to be a good objective of such investigations, the internal variations of shorter than climate change time scales constitute non-linear forcing and are subjected to parametric feed-back (Ji et al, 2014) of which still very little is known. 

Earlier studies of climatological trends in annual mean temperatures over the Indian region showed an increase of about 0.4 $^\circ$C during 1901-82 (e.g., Hingane et al, 1985). Another study pointed to a diurnal asymmetry in this trend with the mean maximum temperatures only showing the warming trend (Rupa Kumar et al, 1994). Using an updated data set Kothawale and Rupa Kumar (2005) have shown that the mean annual temperature over India has increased by 0.05 $^\circ$C/decade for the period 1901-2003 but without any significant diurnal asymmetry during relatively recent years of 1971-2003.

The main purpose of the present study is to partially fill up the global data gap regions by analysing the global warming trend over the Indian region using the gridded data set for the period 1970-2009 and explore the parametric impacts by more basic internal or natural causes to improve the scientific understanding of the integrated phenomena of climatological variation. 
 
\section{Method of Analysis}
The surface-air temperature data over the Indian landmass region at 1$^\circ$x1$^\circ$ latitude-longitude grid points is used in the present analysis for the period 1970-2009. The gridded data set has been created by subjecting the observations to rigorous quality checks, analysis and validation to represent accurate gridded daily mean values of temperature for each grid for all days of 40 years. (Srivastava et al, 2009). This validated dataset was procured from the India Meteorological Department (IMD), Government of India for the current study. Each tabulated data file consists of the values of daily mean surface temperatures over the 1ox1o latitude-longitude grids covering only the Indian land mass areas for each of the days of a particular year. So the 40 years of such data are contained in 40 data files for further processing and analysis. The geographic distribution of these observations is largely homogeneous except in the data sparse regions of Jammu and Kashmir and the north eastern sector of Indian peninsula. Hence, the grid point values pertaining to these regions are less accurate than the rest of the Indian region. 
\begin{figure}[h]
	\begin{center}
		\includegraphics[width=\textwidth]{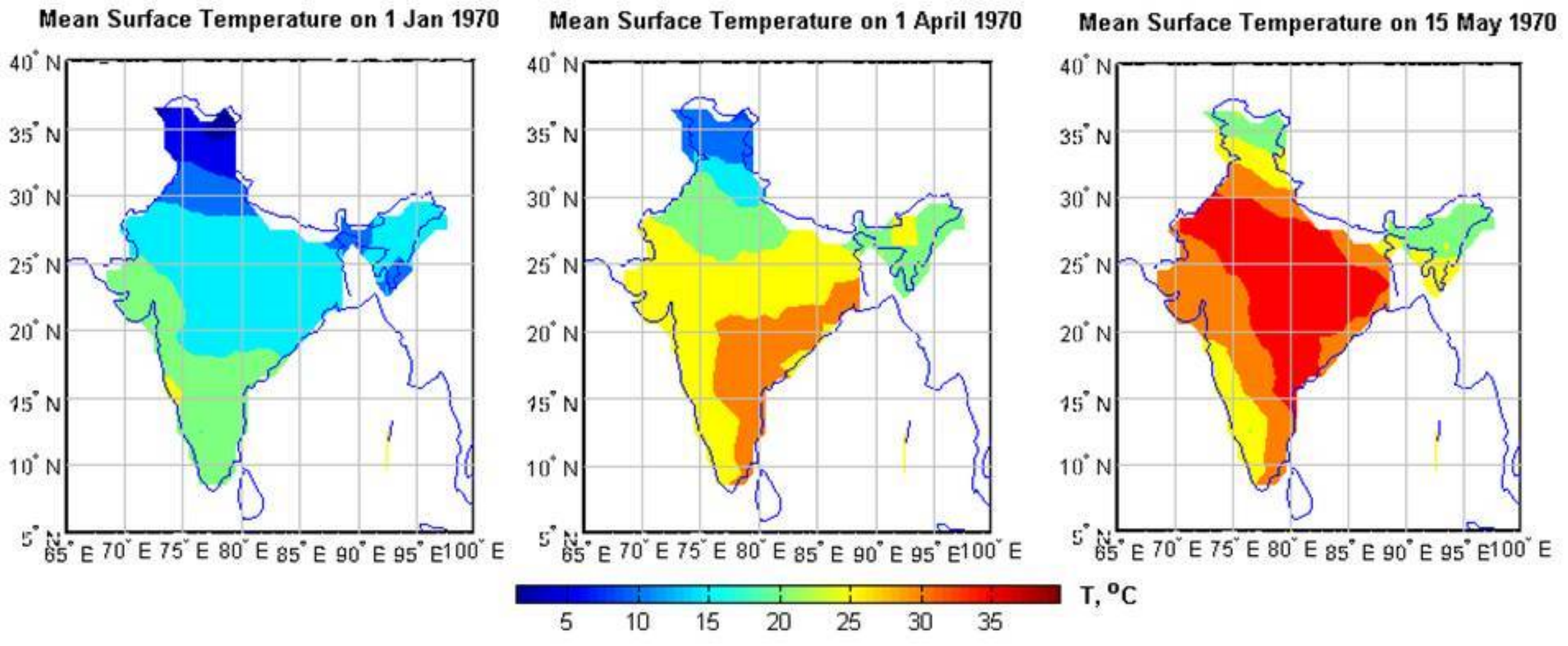}
		\caption{Sample daily mean temperature maps for one day each during winter, equinox and summer months of 1970. The contours have been derived by using the 1$^\circ$x1$^\circ$ grid point data and bilinear interpolation technique for smoothing}
	\end{center}
\end{figure}

The daily mean temperature values for all the available 364(5) days in a particular year are analysed to compute the annual means and standard deviations for each grid point. This analysis is then repeated for all the years from 1970 to 2009. These 40 sets of files are used for further statistical treatment. The main steps include the computations of (a) latitudinal, annual and monthly mean temperatures, (b) 5$^\circ$x5$^\circ$ latitude-longitude average annual temperatures for selected grids of central India and (c) 1$^\circ$x1$^\circ$ pixel level annual mean temperatures covering the whole of India. The yearly series of mean temperatures with different groupings are analysed and compared to make an assessment of any climate change signal and also short period inter annual variations. For retaining the variations due to the increasing linear trend with time as well as any internal variations within this time frame, the absolute values of mean temperatures have been used for the study For understanding possible long term forcing, the annual mean temperature series plots for the entire region are compared with mean yearly series plots of the sunspot numbers, Galactic Cosmic Rays (GCR) neutron fluxes and El Nino and Southern Oscillation (ENSO) indices, the data of which are available in public domains from World or Regional Data Centres. A comprehensive statistical error analysis is carried out to know the bounds of the variances at 95 percent confidence levels pertaining to one sigma excursions in order to qualify the applicability of the observed trends in temperature variations.  Correlation analyses are also carried out to (a) determine the relative sensitivity of different latitude-longitude sectors of Indian region to overall regional warming trend and (b) explore the possible parametric association of variations in global phenomena like the solar activity, GCR neutron flux received on the ground and ENSO indices.

\section{Results and Discussions}
The basic data used in the present study mainly deals with the Daily Mean Temperature (abbreviated as DMT and defined as the average of the daily maximum and minimum temperatures) values provided over the Indian land region for each 1$^\circ$x1$^\circ$ grid point of latitude and longitude. Fig. 1 shows samples of this DMT overlaid onto the Indian maps at 1$^\circ$x1$^\circ$ grids for 3 selected days of winter, equinox and summer months of 1970. The colour code ranges of the 3 maps are kept exactly same for comparison of relative values of DMT. The smoothing for the continuity of the colours is the result of bi-linear interpolation used while drawing the contour diagrams. The progressive seasonal increase in DMT from winter to summer month is clearly seen almost in all parts of the country. 

\begin{figure}[t]
	\begin{center}
		\includegraphics[height=10cm, width=12cm]{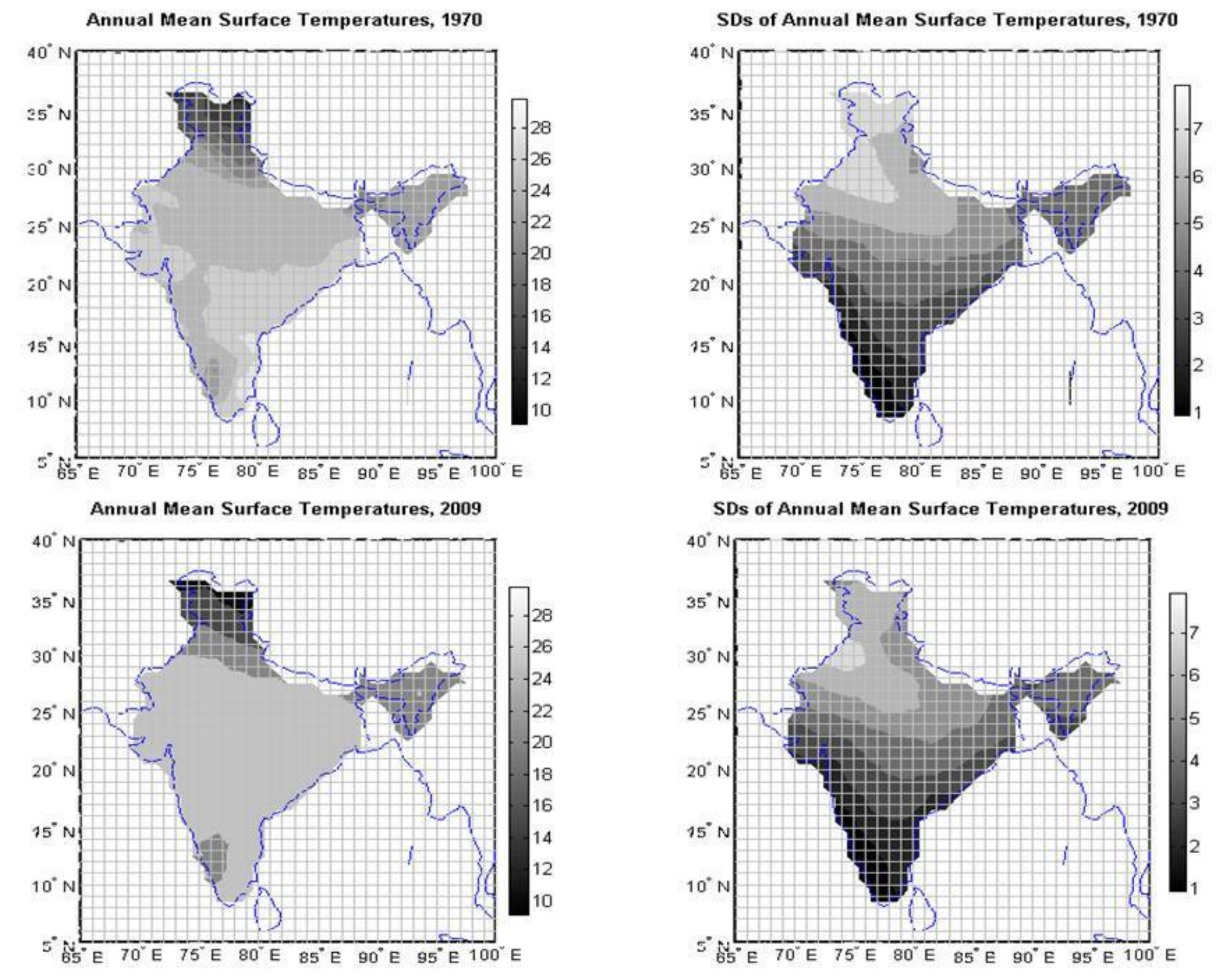}
		\caption{Annual mean temperatures ($^\circ$C) at each 1$^\circ$x1$^\circ$ latitude-longitude grids along with corresponding standard deviations of daily mean temperatures from annual mean temperatures for two years of 1970 and 2009}
	\end{center}
\end{figure}

Fig. 2 shows the Annual Mean Temperatures (AMT) derived from the 364(5) daily values of DMT and the Annual Standard Deviations (ASD) for each 1$^\circ$x1$^\circ$ grid location during the starting and closing years, i.e., 1970 \& 2009 of the data set. A comparison of AMT values during the first and the last year indicates a clear rise in temperature over the whole Indian region from 1970 to 2009. It can be noted that the ASDs increase above 20$^\circ$-25$^\circ$ N latitudes due to large winter-summer temperature gradients or large seasonal variations of DMT compared to the regions closer to the equator. The gradual variation of the seasonal pattern of AMTs along different latitudes is more clearly seen as monthly mean plots in Fig. 3. In this figure the monthly mean temperature profiles averaged along different latitudes at one degree resolution are shown for 1970. The corresponding plot of latitudinal mean DMT-time contours for the same year is shown in Fig. 4. Except for the areas above 30 $^\circ$N, the whole of India experiences very warm weather during most of the days of a year. Using the annual time series temperature data, the analysis is extended to compute the monthly mean latitudinal DMTs for different years. The result of this analysis is presented in Fig. 5 as contour diagrams for each month separately. The long term climatological variations can be discerned from this figure. While the temperature time series for mean monthly DMT along each latitude arc indicates marginally increasing trends (which will be clearer later in this paper), there are some definite inter (multi) annual patterns embedded into the long-term trends. This is clearly shown in a bunch of linear plots given in Fig. 6. Here we can notice the presence of undulations or periodic warming and cooling at intervals of 3-5 years almost synchronously for all latitudinal average series. The simultaneous occurrence of these wave-like structures indicates that the source should be non-local in nature, or global in perspective. The thick black series in this figure is the mean of all latitude series that also shows these oscillatory features pointing to certain phase coherence akin to a source function encompassing a larger region than the Indian region under investigation. For enabling a closer scrutiny of these variations we plot only 3 selected series and the average latitude series in Fig. 7. While there is general agreement on the coherent nature of the patterns in mean temperature series pertaining to 10$^\circ$, 15$^\circ$ and 25$^\circ$ N latitude, there are minor differences. The figure also shows the linear trends of warming present in all the 4 series with a rise in average temperature of ~0.4 $^\circ$C from 1970 to 2009.

\begin{figure}[t]
	\begin{center}
		\includegraphics[width=\textwidth]{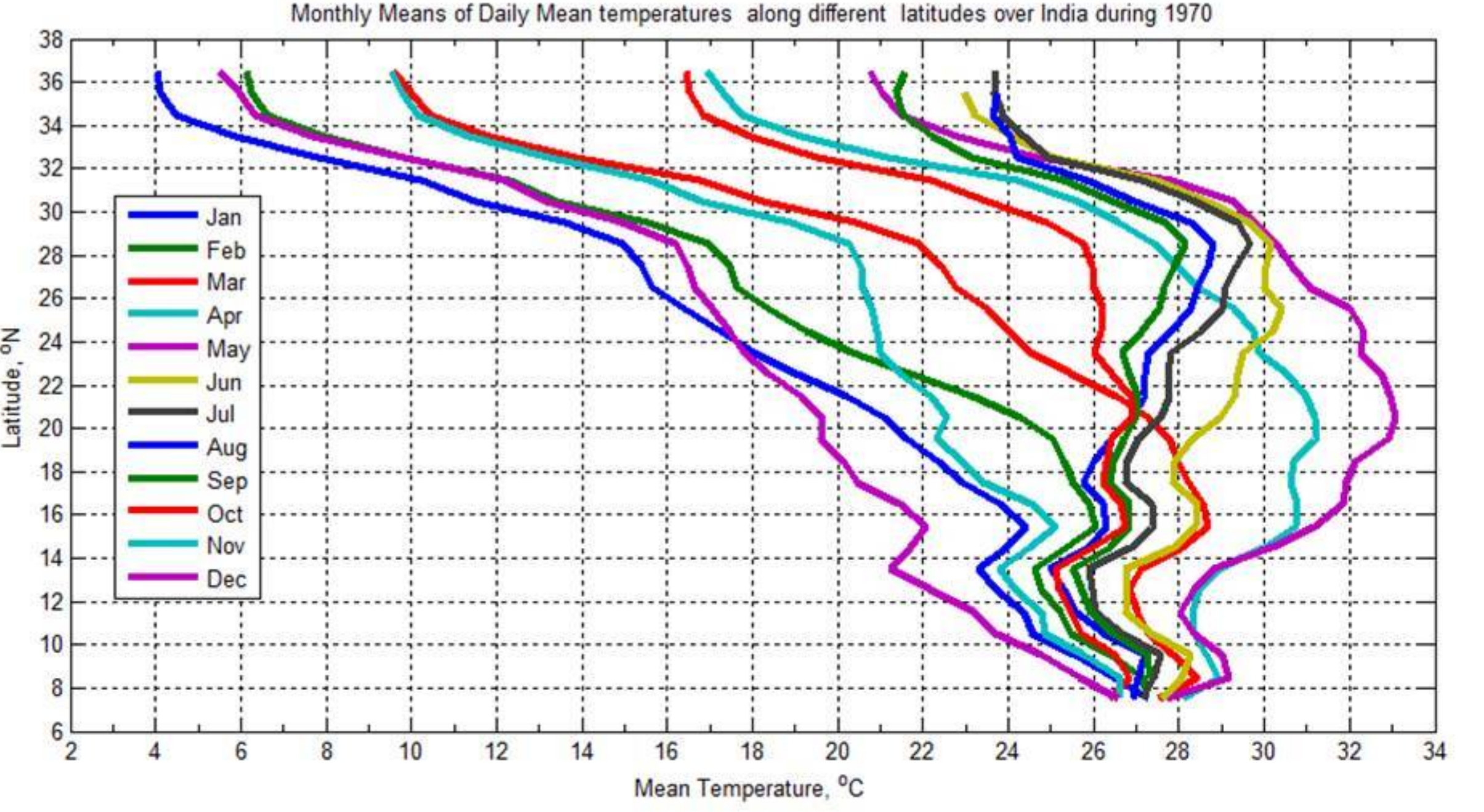}
		\caption{Monthly mean temperature profiles for different latitudes at 1$^\circ$ resolution over the Indian region}
	\end{center}
\end{figure}

\begin{figure}
	\begin{center}
		\includegraphics[height=8cm, width=0.8\paperwidth]{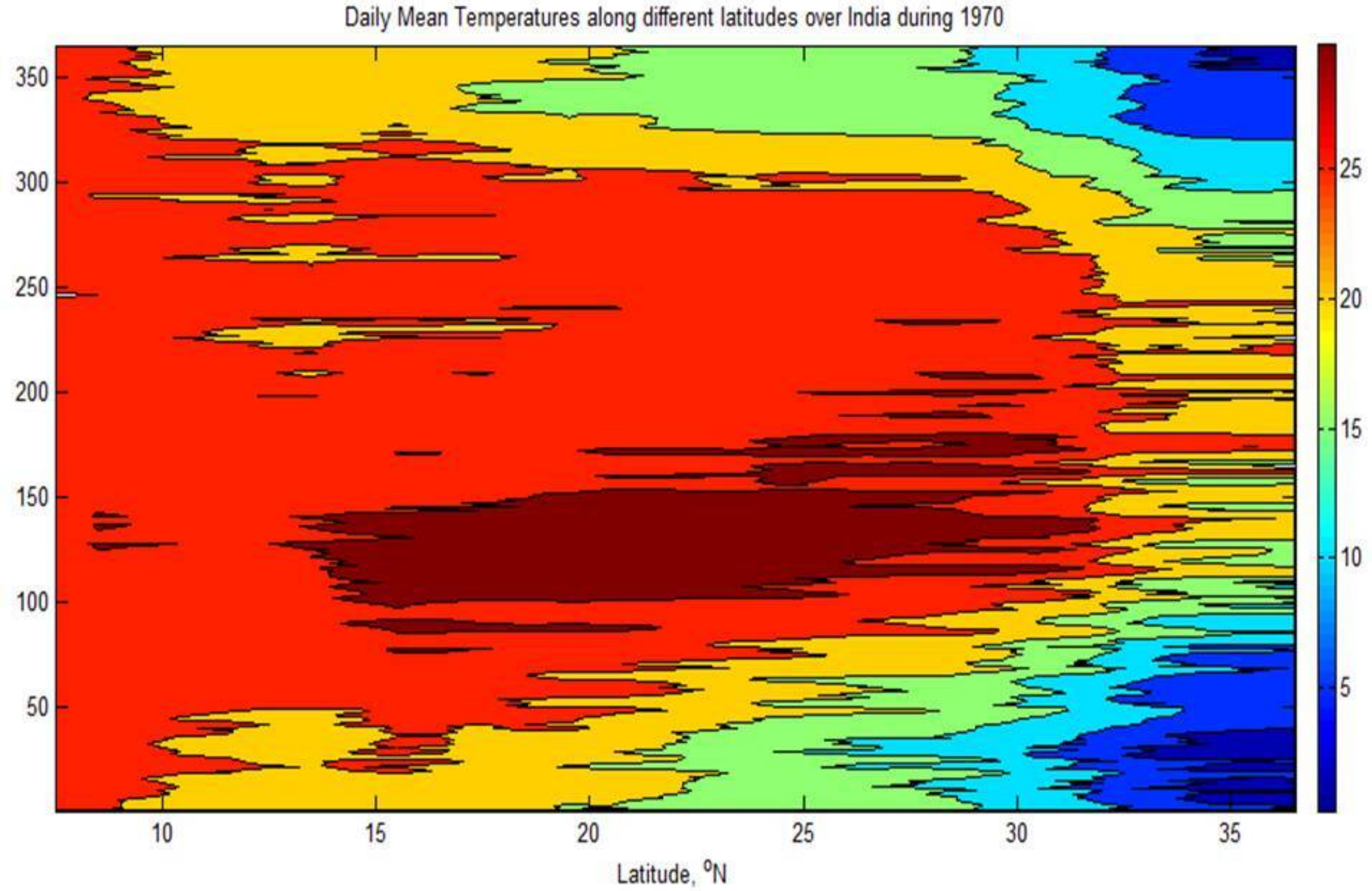}
		\caption{Contour diagram of latitudinal means of daily mean temperatures ($^\circ$C) during 1970 over Indian region}
	\end{center}
\end{figure}

\begin{figure}
	\begin{center}
		\includegraphics[width=\textwidth]{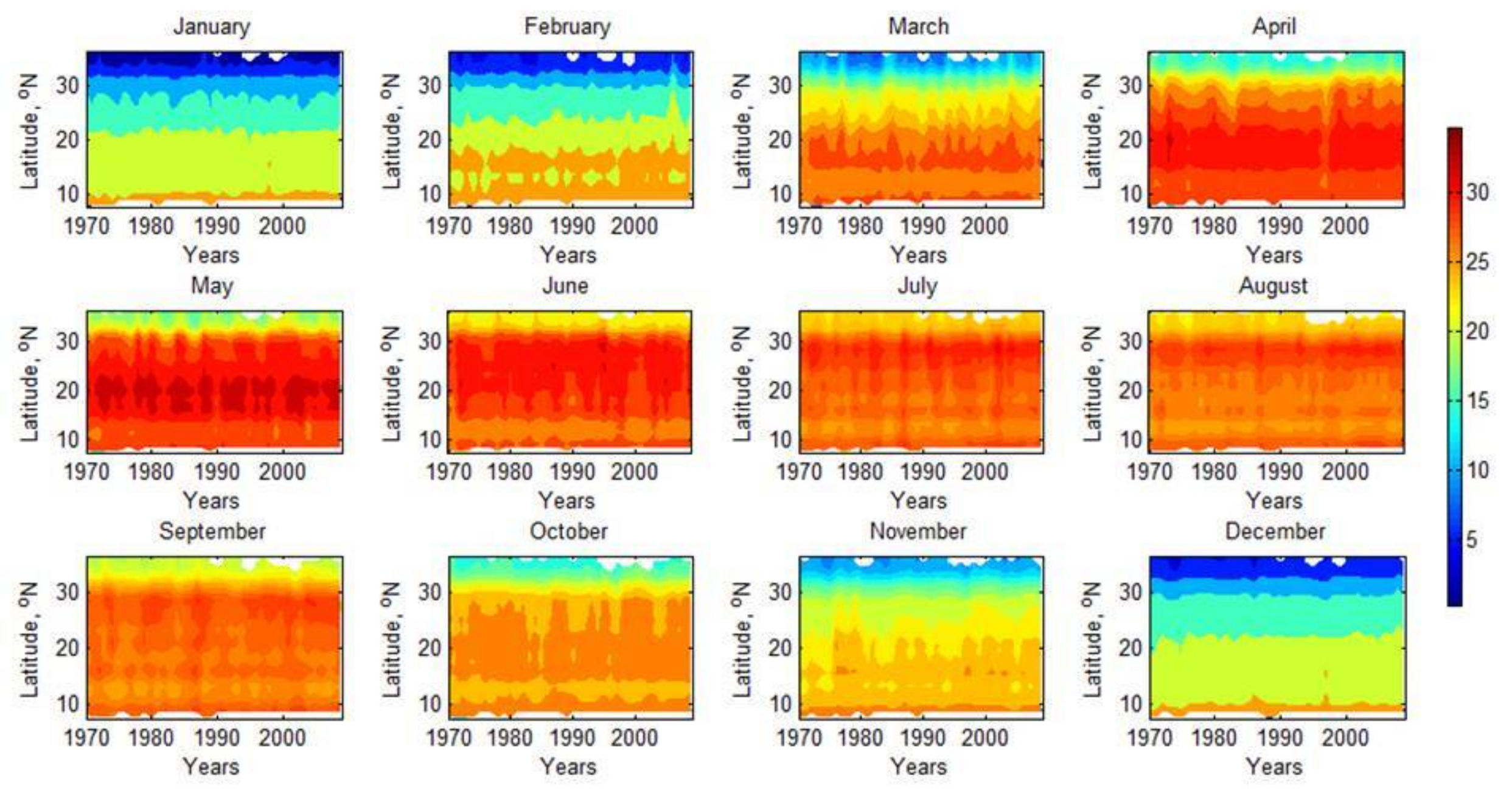}
		\caption{Seasonal patterns of daily mean temperatures ($^\circ$C) profiles (with latitudes) plotted as contours for the time series 1970-2009}
	\end{center}
\end{figure}

Continuing the latitudinal mean temperature analysis, in Fig. 8 we plot the monthly mean temperatures series of DMT averaged over all latitudes. Here also we see similar short period (3-5 years) structures for all the months with a little loss of coherence at times possibly due to large seasonal variations along a longitude line compared to those figures dealing with along the latitude averages. The bold black curve representing the mean of all months (annual) shows similar structures and increasing temperature trend of comparable magnitudes. 

\begin{figure}
	\begin{center}
		\includegraphics[height=8cm, width=15cm]{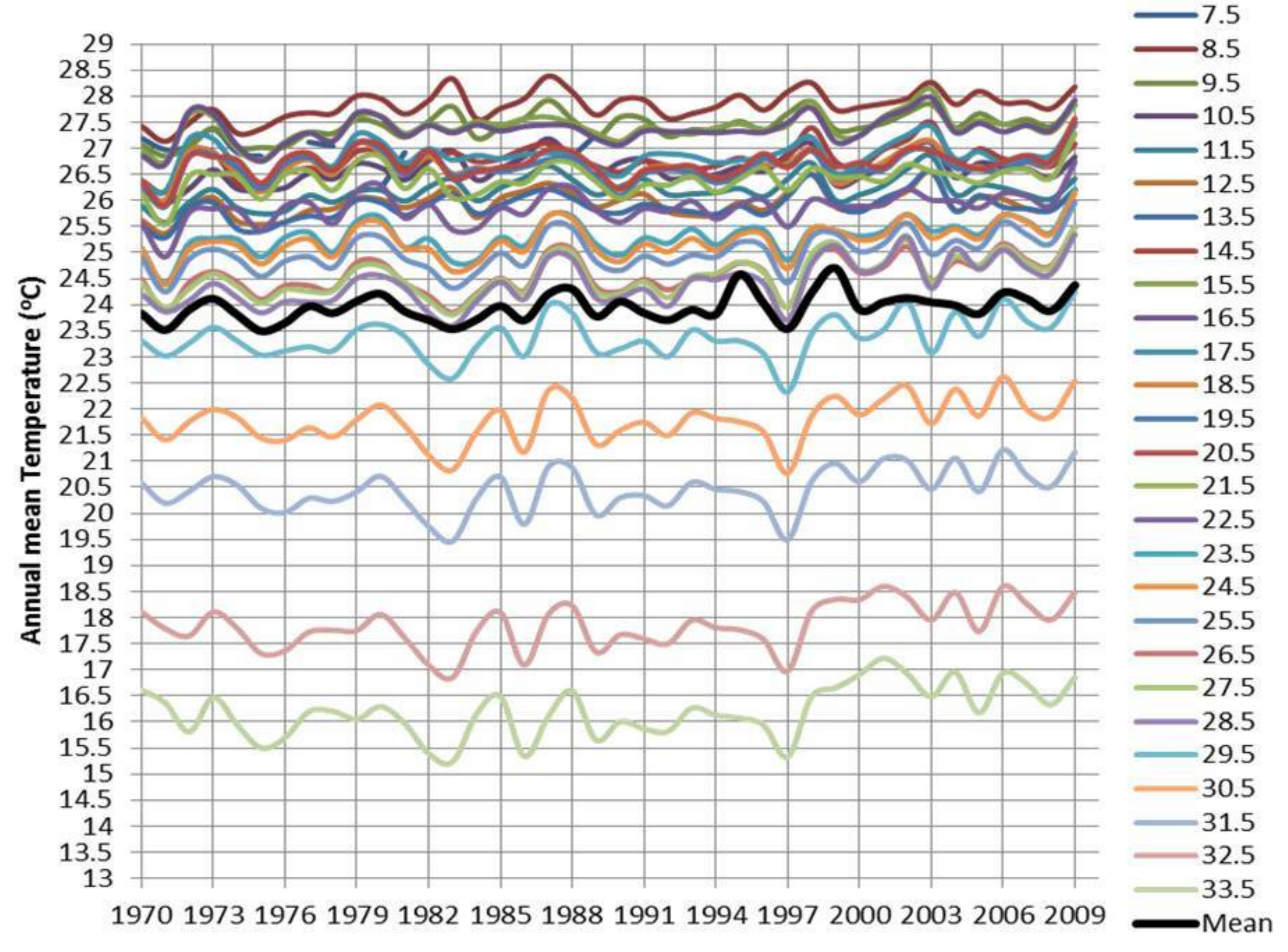}
		\caption{The time series of annual mean temperatures plotted for individual latitude averages over India}
	\end{center}
\end{figure}

\begin{figure}
	\begin{center}
		\includegraphics[width=0.8\paperwidth]{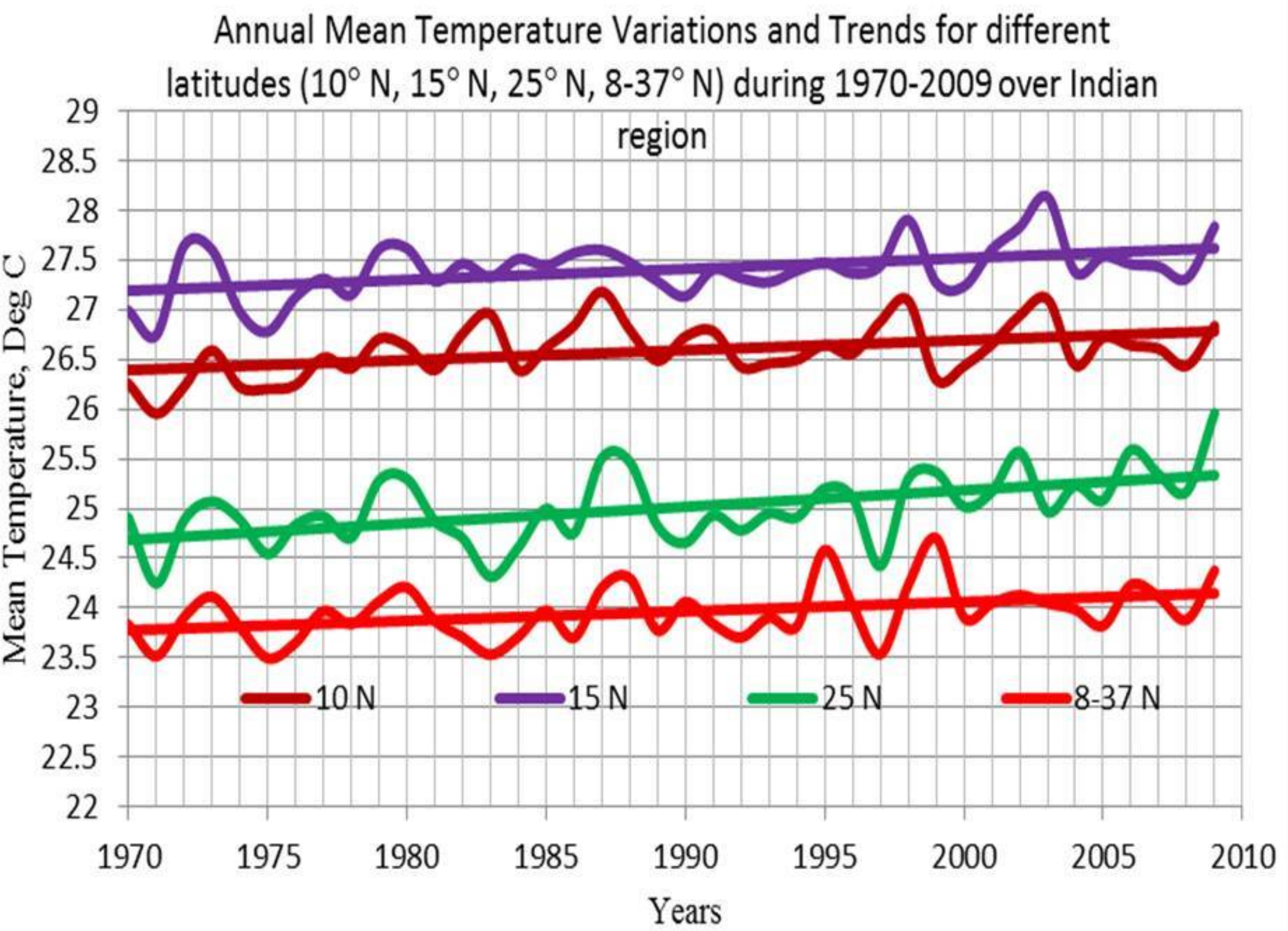}
		\caption{Trends in annual mean temperature series for selected latitude means and for all latitudes representing whole of India divided into latitudes at one degree resolution}
	\end{center}
\end{figure}

\begin{figure}[h]
	\begin{center}
		\includegraphics[width=0.8\paperwidth]{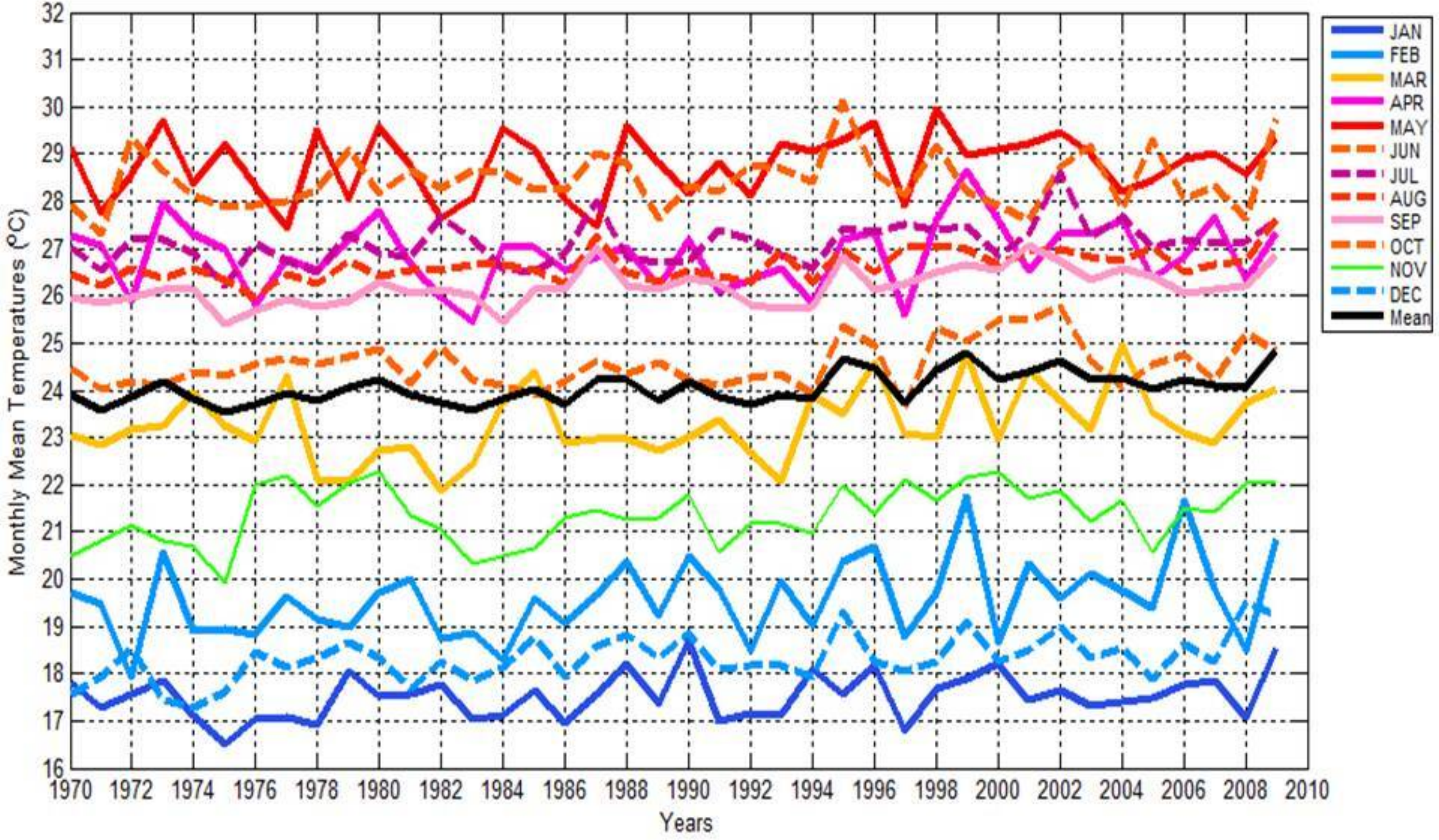}
		\caption{Monthly mean of daily mean temperatures for individual years (by averaging the mean values along all latitudes at one degree interval}
	\end{center}
\end{figure}

The next step to pursue the analysis is to deal with the data at pixel level. For this purpose a 5$^\circ$x5$^\circ$ latitude-longitude grid with a 1$^\circ$x1$^\circ$ grid resolution is selected between 20.5$^\circ$-24.5 $^\circ$N; 75.5$^\circ$-79.5 $^\circ$E with 25 individual pixels. The annual mean of DMT is computed for each pixel each year and averaged for the 5$^\circ$x5$^\circ$ cell. Fig. 9 shows the time series plots of yearly mean temperatures for the whole grid as well as for the points at four corners of this grid (as illustrated) out of 25 sub-grids of 1$^\circ$x1$^\circ$. The standard error bars shown in the 5$^\circ$x5$^\circ$ mean plot are computed by using all the 25 time series at 1$^\circ$x1$^\circ$. While the short period structures and the linear trends are similar, coherence is better between the same latitude pairs indicated by same colour codes. While the outcome of a detailed error analysis is given later in this paper, the errors here are generally lower than the overall long term warming signal of ~0.4 $^\circ$C.

\begin{figure}
	\begin{center}
		\includegraphics[width=0.6\paperwidth]{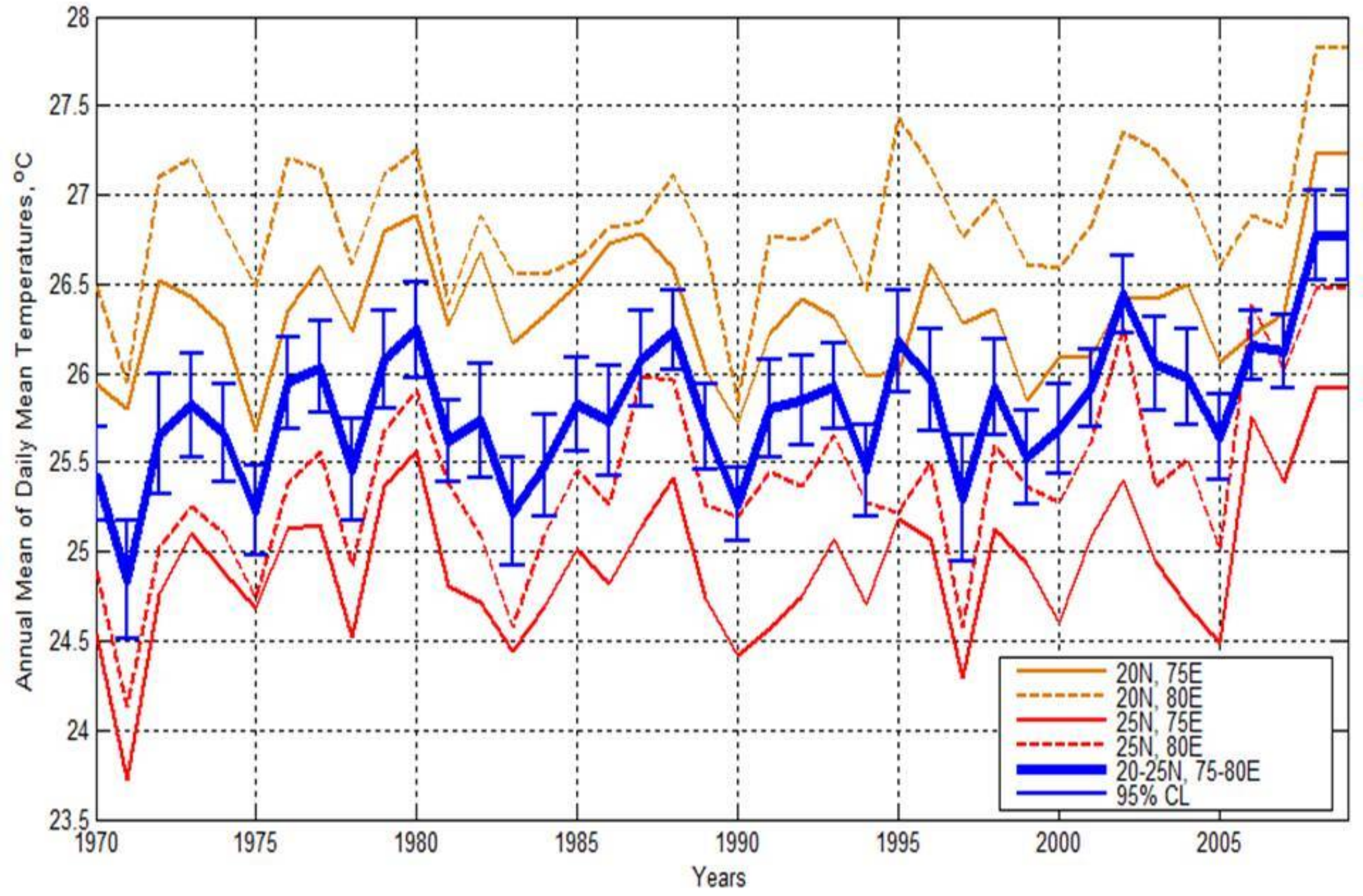}
		\caption{Pixel means of annual temperatures at the centre and 4 corner points of 5$^\circ$x5$^\circ$ grid in central India}
	\end{center}
\end{figure}

\begin{figure}
	\begin{center}
		\includegraphics[width=0.6\paperwidth]{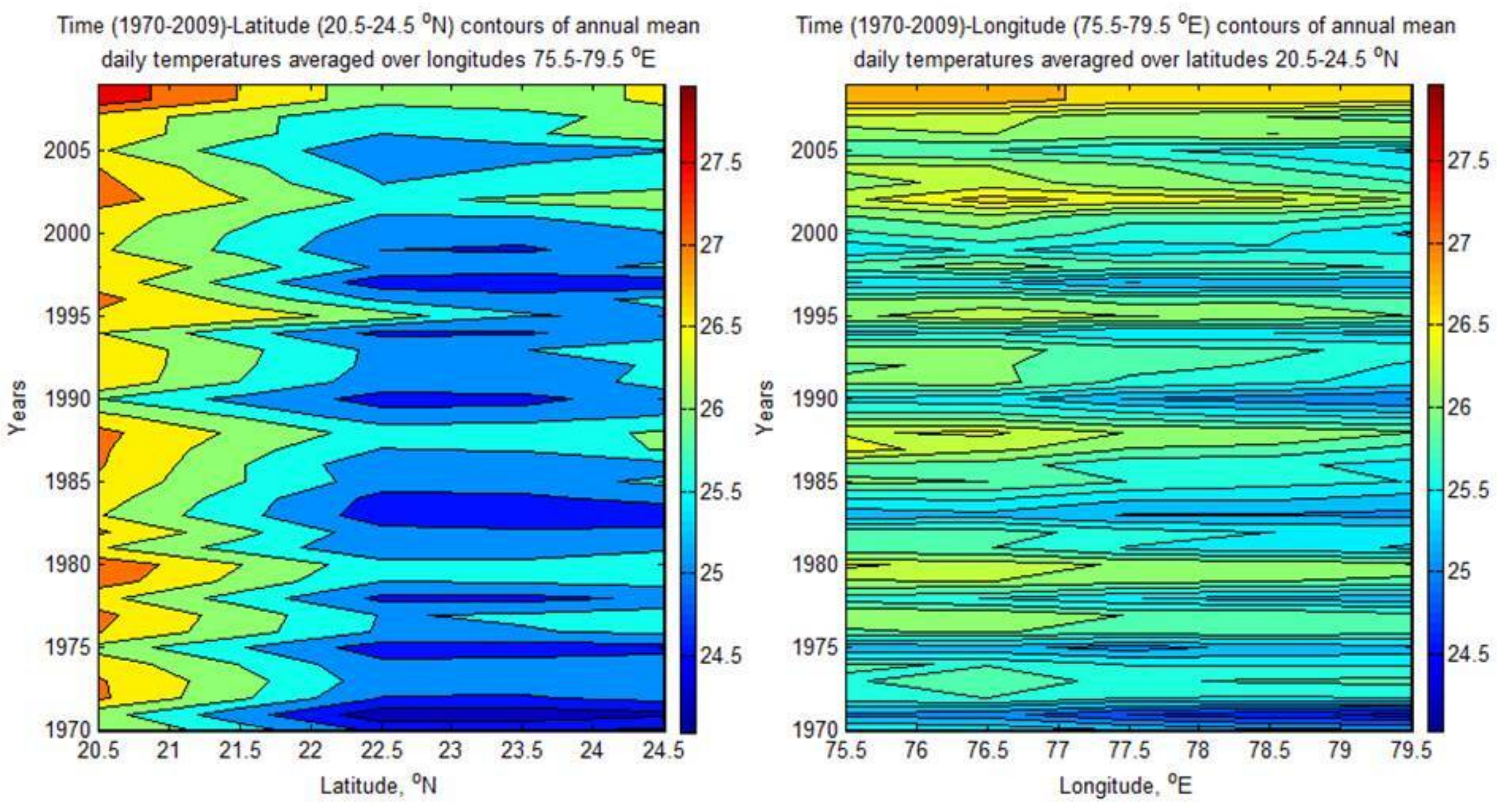}
		\caption{Contours of annual mean temperatures ($^\circ$C) for a) Time (1970-2009)-latitude (20.5-24.5$^\circ$N) and b) Time (1970-2009)-longitude (75.5-79.5 $^\circ$E) }
	\end{center}
\end{figure}

Using all the 25 pixel annual mean temperature values, contour diagrams are generated by taking means along latitudes and longitudes. These plots are shown in Fig. 10. Here the short period structures as well as the increasing trend of mean temperature are clearly seen in both the contour diagrams. Also it is noticed that within 5$^\circ$ of latitude the annual mean temperatures vary more drastically than within 5$^\circ$ of longitude. While it is inferred that the latitudinal temperature gradient essentially depends on mean solar angle, the decreasing temperatures along longitudes towards east indicates a coastal cooling effect due possibly to proximity to the oceanic system.

\begin{figure}
	\begin{center}
		\includegraphics[width=0.6\paperwidth]{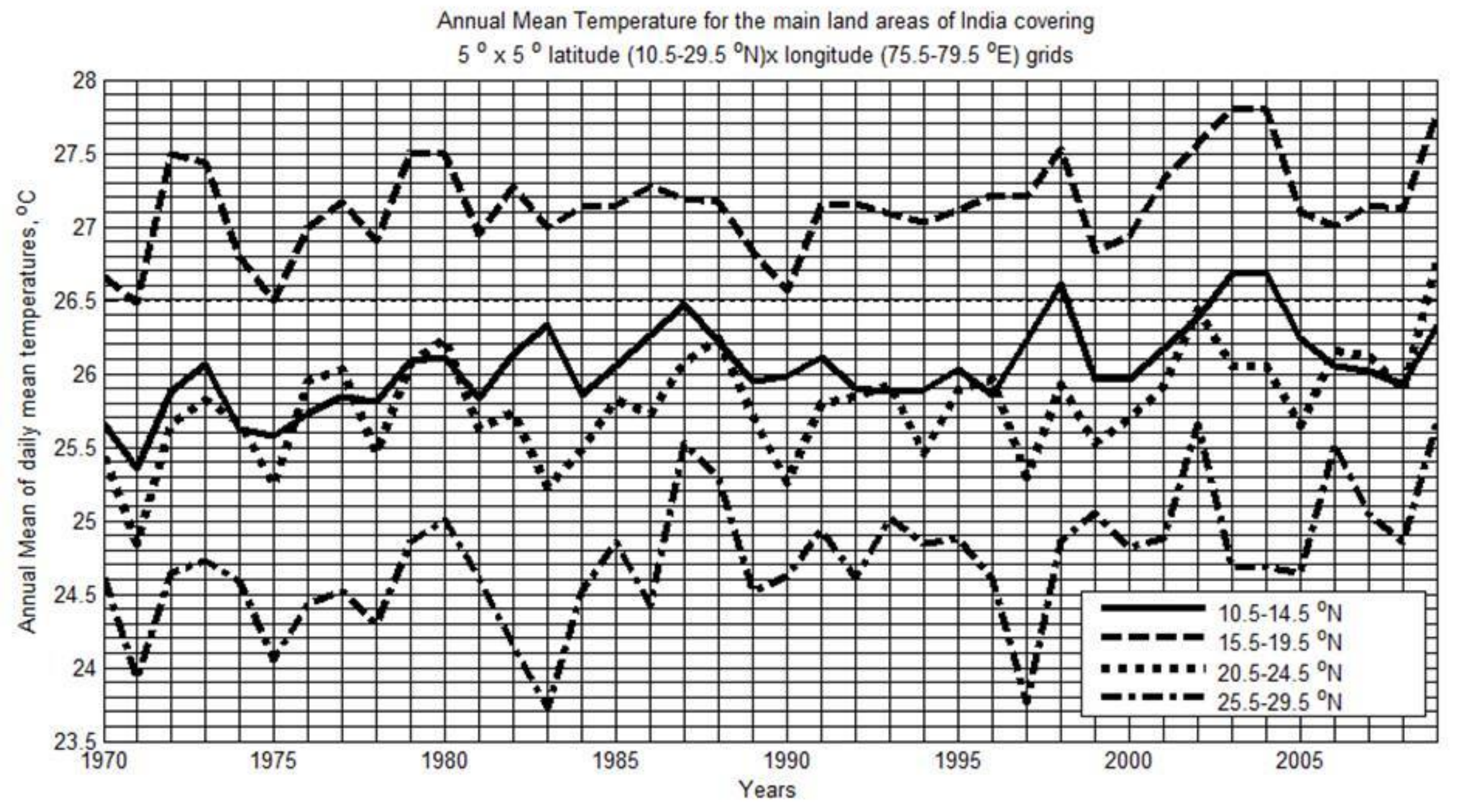}
		\caption{Annual mean temperatures for 4 profiles of 5$^\circ$x5$^\circ$ grids between 10.5-29.5$^\circ$N for the same longitude of 75.5-79.5 $^\circ$E}
	\end{center}
\end{figure}

\begin{figure}
	\begin{center}
		\includegraphics[width=0.6\paperwidth]{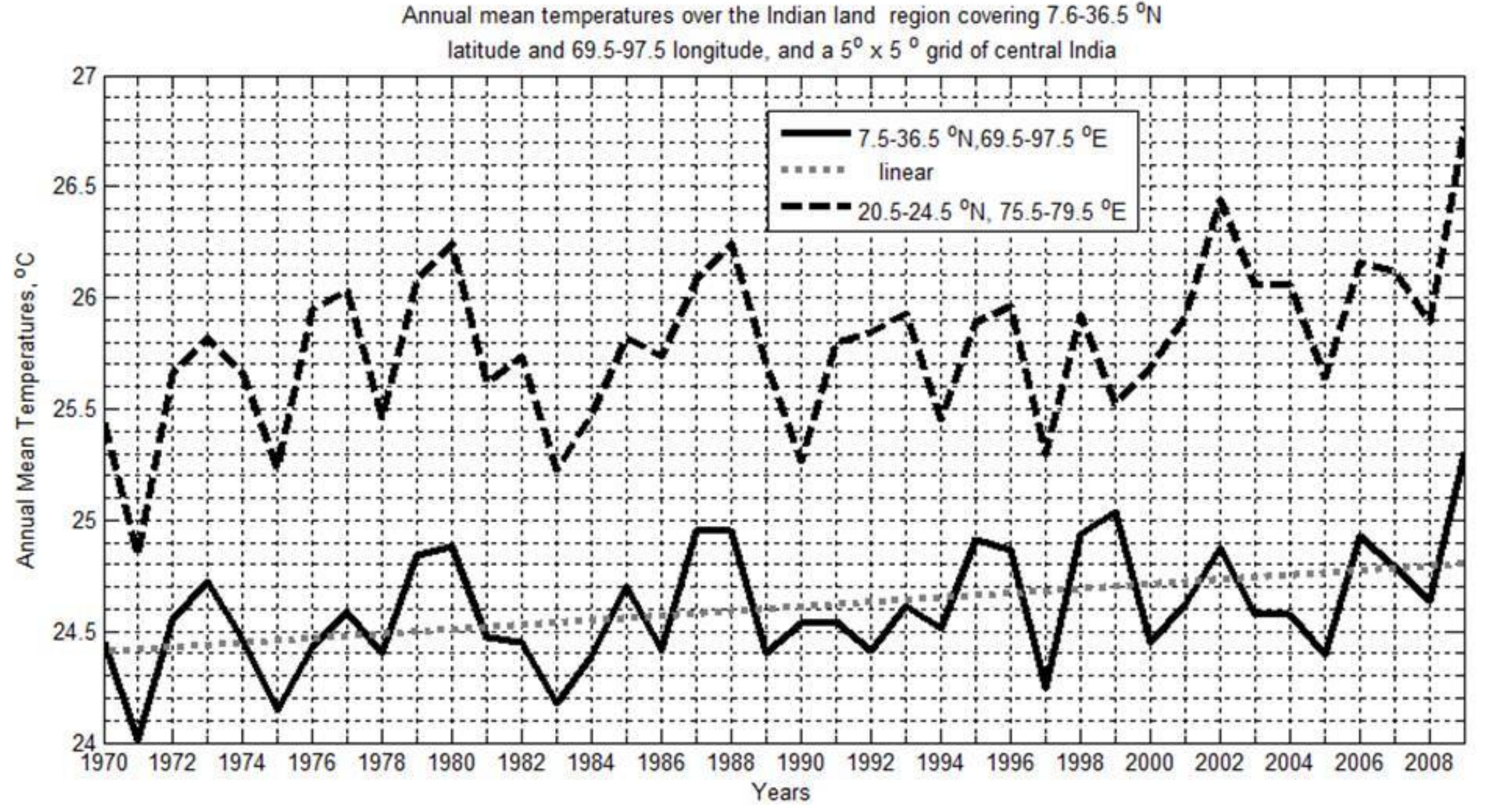}
		\caption{Annual mean temperatures for whole of Indian land area by averaging all the 1$^\circ$x1$^\circ$ grid pixels along with the same parameter estimated for a 5$^\circ$x5$^\circ$ grid of central India}
	\end{center}
\end{figure}

\begin{figure}
	\begin{center}
		\includegraphics[width=0.6\paperwidth]{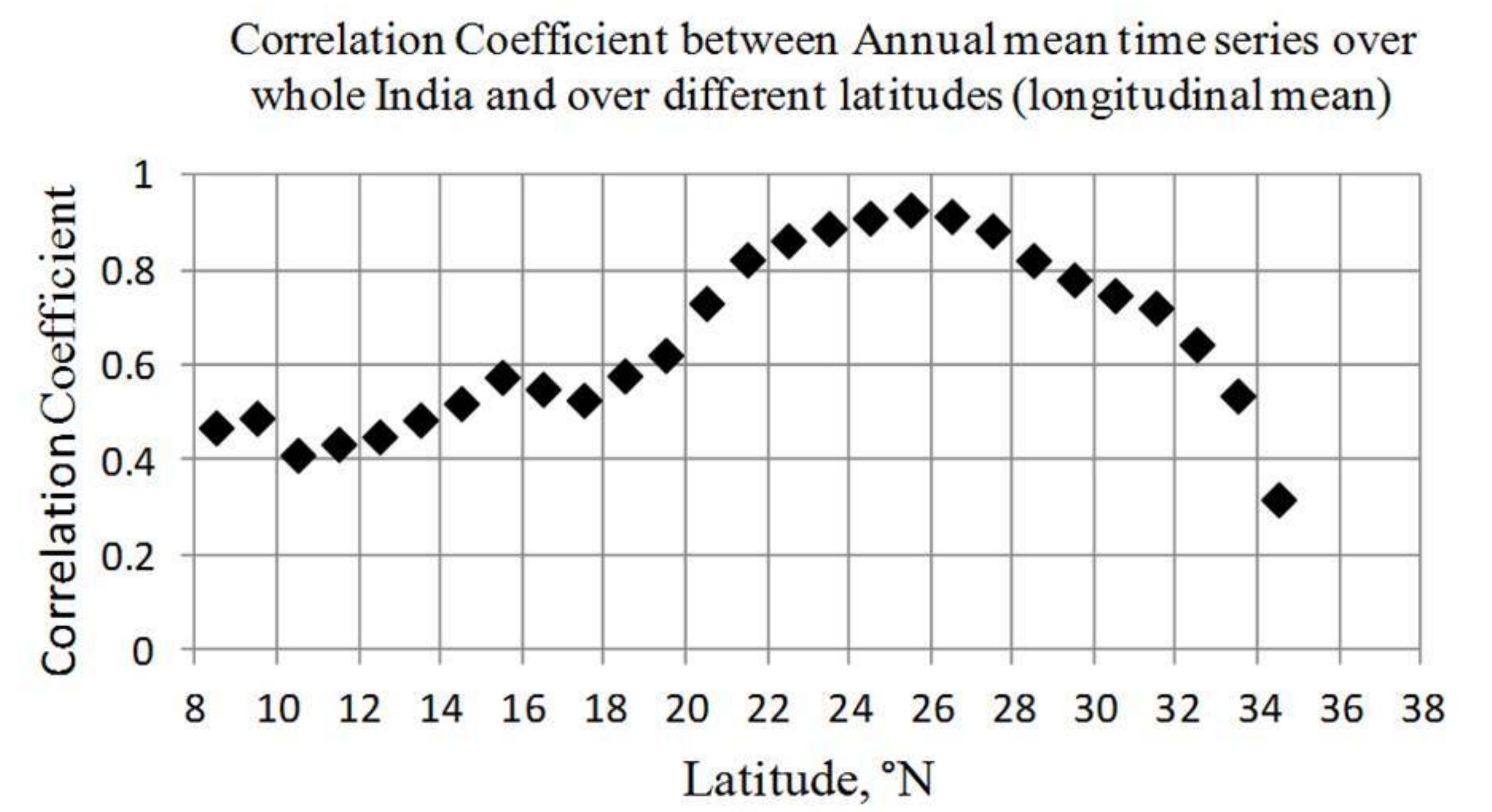}
		\caption{Variation of correlation between the wave-like patterns present in the annual mean temperature series for whole India and that covering longitudinal mean temperatures for different latitude arcs of India}
	\end{center}
\end{figure}

To determine the effect of latitudinal variation over the Indian region on the annual mean temperature sequence at pixel level, 4 sets of 5$^\circ$x5$^\circ$ grids are selected between 10.5-29.5 $^\circ$N for the same longitude range between 75.5-79.5 $^\circ$E. The results are summarised in Fig. 11. All the four curves representing almost whole of India (latitude-wise) corroborate earlier results in terms of the presence of short period structures along with a linear trend of increasing temperature supporting a warming signal due to climate change. By and large the structures are synchronous in time barring a few deviations of minor nature. If we consider 2 pairs of the curves between 10.5-19.5 $^\circ$N latitudes and 20.5-29.5 $^\circ$N, then the wave-like patterns match very well. Further, the yearly time series representing whole of Indian land regions is analysed to obtain the annual mean temperatures for each pixel at 1$^\circ$x1$^\circ$ resolution. From this the annual mean temperatures for the whole of Indian region are derived by averaging over all pixel values. Fig. 12 shows the time series of the annual mean temperatures for whole of India. For comparison the figure includes the annual mean temperatures of the 5$^\circ$x5$^\circ$ grid for central India. Both the curves show striking similarity and the linear trend points to a warming signal of ~0.4 $^\circ$C for the period 1970-2009. To check as how other regions of India correlate with the annual time series of mean temperatures we use the values of mean temperatures computed for different latitudes by averaging for longitudes covered by the given latitude arc. The correlation coefficients are computed for each such latitude time series and the result is shown in Fig. 13. It can be seen that in general there is a good correlation, meaning that the undulations are nearly synchronous. However the high correlation coefficients vary between 0.82-0.93 between the latitude span of 21-28 $^\circ$N representing central India. 

\begin{figure}
	\begin{center}
		\includegraphics[width=0.6\paperwidth]{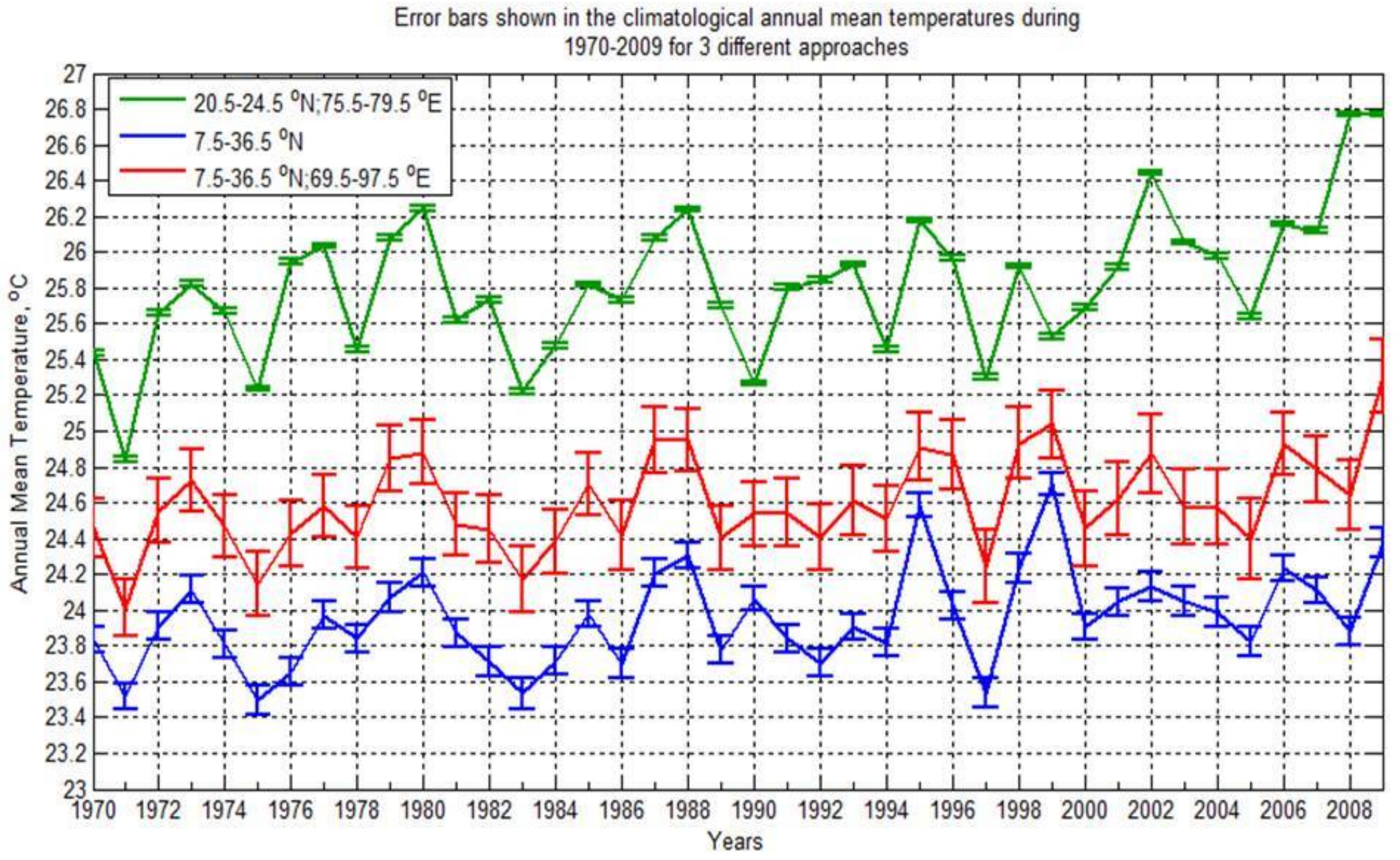}
		\caption{Error bars show the 95\% confidence level values for annual mean temperatures of individual years using 3 different methods (from top to bottom curves): (a) 5$^\circ$x5$^\circ$ latitude-longitude grid in Central India, (b) all 1$^\circ$x1$^\circ$ latitude-longitude grids covering whole of India and (iii) as a function of latitudes at 1$^\circ$ resolution over whole of India (longitudinal mean along latitude arc)}
	\end{center}
\end{figure}
\subsection{Error Analysis}
The data utilised for analysis in this study is mainly intended to determine a statistically sound interpretation of climate change signal over the Indian region. While each grid point data has been considered, the effort has been directed to group the data in different ways to estimate average profiles suitable for inter-comparison. While doing this it is important to determine the relative statistical errors in terms of the confidence limits of the results. Annual average temperatures for each year of the time period of 40 years have been determined in 3 different ways, i.e., for 5$^\circ$x5$^\circ$ grids; along the latitudes at 1$^\circ$ interval; and taking all 1$^\circ$x1$^\circ$ grids into consideration. The standard deviations of daily mean, longitudinal (along latitudes), and latitude-longitude grid mean temperatures are computed for each year and the errors at 95\% confidence level are determined. Fig. 14 gives the relative errors for these 3 different ways of analyses. In all these 3 cases the errors are smaller than the value of the linear warming trend as well as the amplitudes of the oscillatory features.

\subsection{Correlation Analysis}
All the results of above analyses can be divided into two categories, one related to the linear trend of temperature increase attributable to the long term global warming phenomenon and the other showing a periodic variation of temperatures of 3-5 years. Both these characteristics are common in the annual time series of 1970-2009 years for the whole of Indian region irrespective of how the annual mean temperatures are grouped together. Hence like the global cause for the linear warming trend is known to be due to rise in anthropogenic CO2 concentration (Karl and Trenberth, 2003), the ubiquitous short period synchronous patterns in the mean temperatures must be related to some global source function(s). Here we consider variation of 3 parameters during the same time period of 1970-2009 which are known to influence globally. These are (i) Annual mean sunspot numbers as a measure of high energy solar radiation fluxes received by the earth or briefly the solar activity, (ii) Galactic Cosmic Rays (GCR) intensities as revealed by the measurement of neutron counts over McMurdo station in Antarctica and (iii) El Nino and La Nina induced circulation measured by El Nino and Southern Oscillation (ENSO) index or SOI. 

\begin{figure}
	\begin{center}
		\includegraphics[width=0.8\paperwidth]{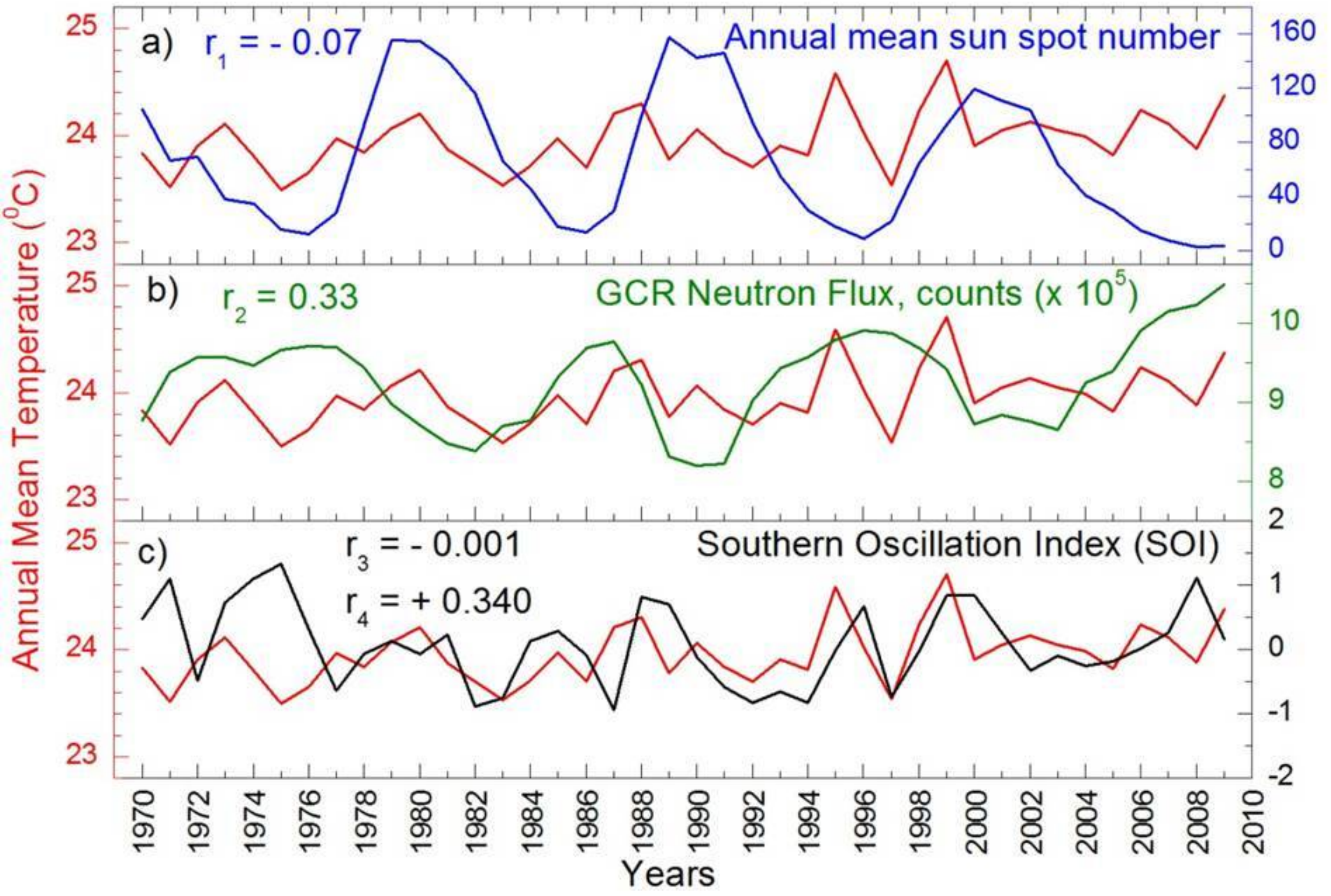}
		\caption{Annual mean temperatures over whole of Indian region plotted with a) mean sunspot numbers (r$_1$=-0.07),  b) mean GCR neutron fluxes (r$_2$=+0.33) and c)  mean Southern Oscillation Indices with r$_3$ (-0.001) and r$_4$ (+0.340) with one year delay in the time series mean temperature for r$_4$}
	\end{center}
\end{figure}

We take the annual mean values of the sunspot numbers, neutron counts and the SOI from open access standard world data centres for the correlation analysis. Fig. 15 shows the results of the 3-plots of these parameters with the annual mean temperatures for the whole India. From the figure, it is clear that GCR has direct positive correlation of +0.33 with the annual mean temperature. In the case of SOI, the correlation coefficient value jumps from 0 to +0.34, if 1-year delayed effect of SOI on the mean temperatures is considered.

\section{Summary and Conclusion}
Global warming studies have been carried out by many groups using annual mean surface-air temperature data of individual stations as well as the data of gridded cells (usually 5$^\circ$x5$^\circ$ in latitude-longitude) at climatological time scales of a few decades to a century. With different data sets and suitable statistical techniques, reasonably consistent results have shown that earth’s mean temperature has gone up by about 0.07 $^\circ$C/decade during the 20th Century. In order to fill up a major gap of such studies, an attempt is made here to conduct a detailed analysis of daily mean temperature data set (at 1$^\circ$x1$^\circ$ resolution in latitude-longitude over land areas of India) during the period 1970-2009, available from IMD. 

The daily mean temperature values are subjected to statistical analysis to derive annual means and standard deviations for spatial coverage at pixel level resolution. Using the time series of annual means derived from different spatial integration, the linear trend of temperatures (considering all 1$^\circ$x1$^\circ$ grids) is found to be about +0.4$^\circ$C during 1970-2009 for the whole India with maximum impact of the warming signal between 20-25$^\circ$N. 

All derived time series of annual mean temperatures show oscillatory features of 3-5 year periodicity. These are present in all pixel, latitude and seasonal variations. The phases of these high frequency structures show striking phase coherence and similarity irrespective of the spatial statistic and the mean amplitudes are of the similar magnitude of \~0.4 $^\circ$C as that of the linear long term trends, but both above the range of errors at 95\% confidence intervals as revealed from the comprehensive error analysis included in this paper. A correlation analysis for the degree of coherence shows that the phases of the short period structures have maximum coherence of about 90\% between 24-26 $^\circ$N latitudes. 

To find any possible relation between the short period temperature structures embedded into the linear trend of annual mean temperatures and other than greenhouse gas sources, e.g. solar activity as determined by sunspot numbers, GCR as obtained from ground based neutron monitors and occurrence of El Nino/La Nina events characterised by the ENSO indices (SOI), a correlation analysis has been carried out. While no significant correlation is found with the solar activity, the results show a positive correlation coefficient of 0.33 with the GCR flux and an improved correlation with SOI from 0 to +0.34 when a time delay of 1-year is introduced in the time series of the mean annual temperatures.

\section*{Acknowledgements}
The authors would like to thank Indian Space Research Organisation (ISRO) for providing financial support under the academic/research project and to IMD for providing the high resolution gridded temperature data over the Indian region.

\section*{References}
\begin{description}
\item Bonfils Safari. Trend analysis of the mean annual temperature in Rwanda during the Last Fifty Two Years, Journal of Environmental Protection, 3, 538-551, 2012

\item Brohan, P., J. J. Kennedy, I. Harris, S. F. B. Tett and P. D. Jones, Uncertainty Estimates in Regional and global observed temperature changes: a new dataset from 1850, Journal of Geophysical Research, 111, D12106. doi:10.1029/2005JD006548, 2006

\item Feng Song, Qi Hu and Weihong Qian, Quality control of daily meteorological data in China, Int. J. Climatol. 24, 853-870, 2004

\item Hingane, L. S., K. Rupa Kumar, and B. V. Ramana Murthy, Long term trends of surface air temperature in India, Int. J. Climatol., 5, 521–528, 1985

\item IPCC Climate Change 2013: The Physical Science Basis (eds Stocker, T. F. et. al) Cambridge Univ Press, 2013.
Ji, F., Wu, Z., Huang, J. \& Chassignet, E. P. Evolution of land surface air temperature trend, Nature Clim. Change 4, 462–466, 2014

\item John Caesar and Lisa Alexander. Large-scale changes in observed daily maximum and minimum temperatures: Creation and analysis of a new gridded data set, Journal of Geophysical Research, 111, D05101, doi:10.1029/2005JD006280, 2006

\item Jones P, Hulme M., The changing temperature of central England in Climates of the British Isles: Present, Past and Future., Hulme M, Barrow E (eds), Routledge: London, 173-196, 1997

\item Jones P. D. and A. Moberg, Hemispheric and large-scale surface air temperature variations: an extensive revision and update to 2001, Journal of Climate, 16, pp. 206-223, 2003, doi:10.1175/1520-0442(2003)016 <0206:HALSSA>2.0.CO;2

\item Kothawale D. R. and K. Rupa Kumar, On the recent changes in surface temperature trends over India, Geophys. Res. Lett., 32, L18714, doi:10.1029/2005GL023528, 2005.

\item Mudelsee M, Climate time series analysis: classical statistical and bootstrap methods, Springer, Dordrecht Heidelberg London New York. 474; Atmospheric and Oceanographic, Sciences Library, 42, 2010

\item Nathaniel W. Chaney, Justin Sheffield, Gabriel Villarini and Eric F. Wood, Development of a high resolution gridded daily meteorological dataset over sub-Saharan Africa: spatial analysis of trends in climate extremes American Met. Soc., 5815-5835, 2014 Doi: 10.1175/JCLI-D-13-00423.1

\item New MG, Hulme M, Jones PD. Representing twentieth-century space-time climate variability, part II: development of 1901–96 monthly grids of terrestrial surface climate. Journal of Climate 13, 2217-2238, 2000

\item Parker D.E. and L.V. Alexander, Global and regional climate in 2001, Weather, 57(9), pp. 328-340. doi:10.1256/00431650260283505, 2002

\item Philip Jones, The reliability of global and hemispheric surface temperature records, Advances in Atmospheric Sciences, 33, pp 1–14, 2016

\item Rupa Kumar, K., K. K. Kumar, and G. B. Pant, Diurnal asymmetry of surface temperature trends over India, Geophys. Res. Lett., 21, 677–680, 1994

\item Srivastava, K., M. Rajeevan and S. R. Kshirsagar, Development of a high resolution daily gridded temperature data set (1969–2005) for the Indian region Atmos. Sci. Let., 2009, DOI: 10.1002/asl.232

\item Thomas R. Karl and Kevin E. Trenberth, Modern global climate change Science, 302, 1719-1723, 2003

\item Timothy J. Osborn, Phil D. Jones and Manoj Joshi, Recent United Kingdom and global temperature variations, Weather, 72(11), 323-329, 2017

\item Yong Lin \& Christian L. E. Franzke, Scale dependency of  the global mean surface temperature trend and its implication for the recent hiatus of global warming Scientific Reports 5:12971, doi: 10.1038/srep12971, 2015
\end{description}

\end{document}